\documentstyle[aps,preprint]{revtex}
\begin{document}
\tighten
\draft
\title{The Detectability of Relic (Squeezed) Gravitational Waves by
Laser Interferometers}

\author{L. P. Grishchuk\thanks{e-mail: grishchuk@astro.cf.ac.uk}}
\address{Department of Physics and Astronomy, Cardiff University, 
Cardiff CF2 3YB, United Kingdom \\ and \\ Sternberg Astronomical
Institute, Moscow University, Moscow 119899, Russia}

\date{\today}
\maketitle

\begin{abstract}
It is shown that the expected amplitudes and specific correlation 
properties of the relic (squeezed) gravitational wave background 
may allow the registration of the relic gravitational waves by 
the first generation of sensitive gravity-wave detectors.  
\end{abstract}


%\end{titlepage}    

Relic gravitational waves are inevitably generated by strong 
variable gravitational field of the very early Universe which 
parametrically (superadiabatically) amplifies the zero-point quantum
oscillations of the gravitational waves [1]. The initial vacuum 
quantum state of each pair of waves 
with oppositely directed momenta evolves into a highly correlated 
multiparticle state known as the two-mode squeezed vacuum quantum 
state [2]. (For a recent review of squeezed states see [3]). In 
cosmological context, the squeezing manifests itself in a specific 
standing-wave pattern and periodic correlation functions of the
generated field [4]. The main point of this paper is to show that this 
signature may significantly facilitate the detection of the 
relic (squeezed) gravitational wave background. It is possible that
the appropriate data processing may allow the detection of relic 
gravitational waves by the forthcoming sensitive instruments, such as 
the initial laser interferometers in LIGO [5], VIRGO [6], GEO600 [7].
\par
Before analyzing the theoretical predictions in light of the 
current experimental situation, it is necessary to explain
the origin of these predictions.
\par
We consider the cosmological gravitational wave field $h_{ij}$       
defined by the expression
\begin{equation}
\label{1}
{\rm d}s^2 = a^2({\eta})[{\rm d}\eta^2 - (\delta_{ij} + h_{ij})
{\rm d}x^i{\rm d}x^j]~.
\end{equation} 
The Heisenberg operator for the quantized real field $h_{ij}$ can 
be written as
\begin{eqnarray}
\label{2}
h_{ij} (\eta ,{\bf x})
= {C\over (2\pi )^{3/2}} \int_{-\infty}^\infty d^3{\bf n}
  \sum_{s=1}^2~{\stackrel{s}{p}}_{ij} ({\bf n})
   {1\over \sqrt{2n}}
\left[ {\stackrel{s}{h}}_n (\eta ) e^{i{\bf nx}}~
                 {\stackrel{s}{c}}_{\bf n}
                +{\stackrel{s}{h}}_n^{\ast}(\eta ) e^{-i{\bf nx}}~
                 {\stackrel{s}{c}}_{\bf n}^{\dag}  \right],
\end{eqnarray}
where $C=\sqrt{16\pi}~l_{Pl}$ and the Planck length is 
$l_{Pl}=(G\hbar /c^3)^{1/2}$, the creation and annihilation operators satisfy 
${\stackrel{s}{c}}_{\bf n}|0\rangle =0$, 
$[{\stackrel{s'}{c}}_{\bf n},~{\stackrel{s}{c}}_{{\bf m}}^{\dag}]=
\delta_{ss'}\delta^3({\bf n}-{\bf m})$, and the wave number $n$ is
related to the wave vector ${\bf n}$ by 
$n = (\delta_{ij}n^in^j)^{1/2}$. 
The two polarisation tensors ${\stackrel{s}{p}}_{ij}({\bf n})$ $(s = 1, 2)$ 
obey the conditions
\[
 {\stackrel{s}{p}}_{ij}n^j = 0, ~~
 {\stackrel{s}{p}}_{ij}\delta^{ij} = 0, ~~
 {\stackrel{s'}{p}}_{ij}
 {\stackrel{s}{p}}~^{ij} = 2\delta_{ss'}, ~~
 {\stackrel{s}{p}}_{ij}(-{\bf n}) = {\stackrel{s}{p}}_{ij}({\bf n}).
\]
For every wave number $n$  
and each polarisation component $s$,
the functions ${\stackrel{s}{h}}_n(\eta )$ have the form 
\begin{equation}
\label{3}
  {\stackrel{s}{h}}_n(\eta ) = {1\over a(\eta )} 
  [{\stackrel{s}{u}}_n(\eta ) + {\stackrel{s}{v}}_n^{\ast} (\eta )],  
\end{equation}
where ${\stackrel{s}{u}}_n(\eta )$ and ${\stackrel{s}{v}}_n(\eta )$
are expressed in terms of the three
real functions: $r_n$ - squeeze parameter, $\phi_n$ - squeeze angle, 
$\theta_n$ - rotation angle,   
\begin{equation}
\label{4}
   u_n = e^{i{\theta}_n} \cosh~{r}_n, \qquad
   v_n = e^{-i({\theta}_n - 2{\phi}_n )} \sinh~{r}_n.
\end{equation}
The functions $r_n(\eta)$, $\phi_n(\eta)$, $\theta_n(\eta)$ are governed by
the dynamical equations [4]: 
\begin{equation}
\label{5}
r_n^{\prime} = \frac{a^{\prime}}{a} \cos{2{\phi}_n}, \quad
\phi_n^{\prime} = -n - \frac{a^{\prime}}{a} \sin{2{\phi}_n}\coth~2{r}_n, \quad 
\theta_n^{\prime} = -n - \frac{a^{\prime}}{a} \sin{2{\phi}_n}\tanh~{r}_n,
\end{equation}
where $^{\prime} = {d}/{d\eta}$, and the evolution begins 
from $r_n = 0$, which
characterizes the initial vacuum state. The dynamical equations and their
solutions are identical for both polarisation components $s$. 
\par
The present day values of $r_n$ and $\phi_n$ 
are essentially all we need to calculate. The mean number of particles
in a two-mode squeezed state is $2\sinh^2{r_n}$ (for each $s$). This number 
determines the 
mean square amplitude of the gravitational wave field. The time behaviour of
the squeeze angle $\phi_n$ determines the time dependence of the correlation 
functions of the field. The amplification (that is, the growth of $r_n$) 
governed 
by (5) is different for different wave numbers $n$. Therefore, 
the present day results depend on the present day frequency $\nu$
($\nu = {cn}/{2 \pi a}$)    
measured in $Hz$.  
\par 
In the short-wavelength (high-frequency) regime, that is,  
during intervals of time
when the wavelength $\lambda(\eta) = 2 \pi a/n$ is shorter than the Hubble 
radius $l(\eta) = a^2/a^{\prime}$,
the term $n$ in (5) is dominant, and the functions $\phi_n(\eta)$,
$\theta_n(\eta)$ are $\phi_n = -n(\eta + const)$, $\theta_n = \phi_n$.  
The factor $\cos 2\phi_n$ is a quickly oscillating
function of time, so the squeeze parameter $r_n$ stays practically constant.
\par
In the opposite, long-wavelength regime, the term $n$ can be 
neglected.
The function $\phi_n$ is $\tan \phi_n(\eta) \approx const/a^2(\eta)$,
and the squeeze angle
quickly approaches one of the two values $\phi_n = 0$ or $\phi_n = \pi$   
(an analog of the ``phase bifurcation" [8]). 
The squeeze parameter $r_n(\eta)$ grows with time according to   
\begin{equation}
\label{6}
r_n(\eta) \approx ln \frac{a(\eta)}{a_*}\quad, 
\end{equation} 
where $a_*$ is the value of $a(\eta)$ when the long-wavelength regime, 
for a given $n$, begins. The final amount of $r_n$ is
\begin{equation}
\label{7}
r_n \approx ln \frac{a_{**}}{a_*}\quad, 
\end{equation} 
where $a_{**}$ is the value of $a(\eta)$ when the long-wavelength regime and 
amplification come to the end. 
\par
After the end of amplification, the accumulated
(and typically large) squeeze parameter $r_n$ stays approximately constant. 
The complex functions 
${\stackrel{s}{u}}_n(\eta ) + {\stackrel{s}{v}}_n^{\ast}(\eta )$
become practically real, and one has 
${\stackrel{s}{h}}_n(\eta ) \approx {\stackrel{s}{h}}_n^{\ast}(\eta ) \approx
\frac{1}{a} e^{r_n} \cos \phi_n(\eta)$.  
Every mode $n$ of the field (2) takes the form of 
a product of a function of time and a (random, operator-valued) 
function of spatial coordinates, that is, the mode acquires a 
standing-wave pattern. The periodic dependence $\cos \phi_n(\eta)$
is a subject of our further inquiry.  
\par 
The numerical results depend on the concrete behaviour of the pump
field represented by the cosmological scale factor $a(\eta)$. We know that 
the present
matter-dominated stage $a(\eta) \propto \eta^2$ was preceeded by the 
radiation-dominated stage $a(\eta) \propto \eta$.  
The function $a(\eta)$ describing the initial stage of expansion
of the very early Universe (before
the era of primordial nucleosynthesis) is not known. It is convenient to
parameterize $a(\eta)$ by power-law functions in terms of $\eta$, since 
they produce gravitational waves with 
power-law spectra in terms of $\nu$~[1].
Concretely, we take $a_i(\eta)$ at the 
initial stage of expansion as
$a_i(\eta) = l_o|\eta|^{1 + \beta}$ 
where $\eta$ grows from $- \infty$,  and $\beta < -1$. From 
$\eta = \eta_1$, $\eta_1 < 0$, 
the initial stage is followed by the radiation-dominated stage
$a_e(\eta) = l_oa_e(\eta - \eta_e)$ and then, from $\eta = \eta_2$,
by the matter-dominated stage 
$a_m(\eta) = l_oa_m(\eta - \eta_m)^2$. 
The constants $a_e$, $a_m$, $\eta_e$, $\eta_m$            
are expressed in terms of the fundamental parameters $l_o$, $\beta$ through
the continuous joining of $a(\eta)$ and $a^{\prime}(\eta)$ at $\eta_1$,
$\eta_2$.
The present era is defined by the observationally 
known value of the Hubble radius 
$l_H = c/H \approx 2 \times 10^{28}~{\rm cm}$.  
We denote this time by $\eta_R$ and choose $\eta_R - \eta_m = 1$,
so that $a(\eta_R) = 2l_H$. The ratio $a(\eta_R)/a(\eta_2) = z$ is believed
to be around $z = 10^4$. Some information about $l_o$ and $\beta$ is 
provided by the data on the microwave background anisotropies [9, 10], 
and we will use it below. 
\par
The known function $a(\eta)$ and equations (5) allow us to find the 
present day values of $r_n$ as a function of $n$. Let the wave numbers
$n_H (n_H = 4 \pi)$, $n_m (n_m = \sqrt z n_H)$, $n_c$
denote the waves which are leaving the long-wavelength 
regime at, correspondingly, $\eta_R$, $\eta_2$, $\eta_1$. 
The $n_c$ is the value of $n$ for which $a_{**} = a_*$. 
The shorter waves, with $n > n_c$, have never been in the 
amplifying long-wavelength regime. (The present day 
frequency $\nu_c$ is around $10^{10}~ Hz$, as we will see below). Thus, 
$r_n = 0$ for $n_c \leq n$, $r_n = ln [(n/n_c)^\beta]$ for 
$n_m \leq n \leq n_c$,
$r_n = ln [(n/n_c)^{\beta - 1}(n_m/n_c)]$ for $n_H \leq n \leq n_m$, and
$r_n = ln[(n/n_c)^{\beta + 1} (n_m/n_H)(n_c/n_H)]$ for $n \leq n_H$.  
The $e^{r_n}$ is much larger than $1$ for all frequencies $n \ll n_c$ 
$(\nu \ll \nu_c)$.          
\par
The mean value of the field $h_{ij}$ is zero,                       
$\langle 0|h_{ij}(\eta, {\bf x})|0\rangle = 0$.
The variance 
\[
\langle 0|h_{ij}(\eta, {\bf x})h^{ij}(\eta, {\bf x})|0\rangle ~\equiv~
\langle h^2 \rangle 
\] 
is not zero, and it determines the mean square amplitude of the generated
field - the quantity of interest for the experiment. Taking the product
of two expressions (2), one can show that   	 
\begin{equation}
\label{8}
\langle h^2 \rangle = \frac{C^2}{2\pi^2} \int_0^\infty n \sum_{s=1}^2
\Big| {\stackrel{s}{h}}_n(\eta )\Big|^2 ~{\rm d}n = 
\frac{C^2}{{\pi^2}a^2 } \int_0^\infty n {\rm d}n (\cosh2{r_n} + 
\cos2{\phi_n}\sinh2{r_n}). 
\end{equation}
Eq. (8) can also be written as
\[
\langle h^2 \rangle = \int_0^\infty h^2(n, \eta) \frac{{\rm d}n}{n}, 
\] 
where, for the present era, 
\begin{equation}
\label{9}
h(n, \eta)\approx \frac{C}{\pi}\frac{1}{a(\eta_R)}n e^{r_n}\cos\phi_n(\eta)~.
\end{equation}
The further reduction of this formula gives
\begin{equation}
\label{10}
h(n) \approx  A \left(n\over n_H\right)^{\beta+2}~~,~~~~~  n\leq n_H, 
\end{equation}
\begin{equation}
\label{11}
h(n,\eta) \approx A \cos\phi_n (\eta) \left(n\over n_H\right)^\beta~~, 
~~~~~ n_H \leq n\leq n_m	, 
\end{equation}
\begin{equation}
\label{12}
h(n,\eta) \approx A \cos\phi_n (\eta) \left(n\over n_H\right)^{\beta+1}  
\left(n_H\over n_m \right)~~,~~~~~  n_m \leq n\leq n_c, 
\end{equation}
where 
\[
A = \frac{l_{Pl}}{l_o}\frac{8\sqrt{\pi}2^{\beta+2}}{|1+\beta|^{\beta+1}}~.
\]
\par 
The available information on the microwave background 
anisotropies [9, 10] allows us to determine the parameters $A$ and $\beta$.
The quadrupole anisotropy produced by the spectrum (10) - (12) is
mainly accounted for by the wave numbers near $n_H$. Thus, 
the numerical value of the quadrupole anisotropy produced by relic 
gravitational waves is approximately equal to $A$. 
Since (according to [11]) the quadrupole contribution of relic 
gravitational waves is not smaller than that produced by primordial 
density perturbations,
this gives us $A\approx 10^{-5}$. We do not know experimentally whether
a significant part of the quadrupole signal is indeed provided by relic 
gravitational waves,
but we can at least assume this. The evaluation of the spectral 
index n of the primordial perturbations resulted 
in ${\rm n} = 1.2 \pm 0.3$ [10] or even in a significantly higher 
value ${\rm n} = 1.84 \pm 0.29$ [12] (see also [19], where one of the best
fits corresponds to ${\rm n} = 1.4$). We interpret [13] these evaluations 
as an indication that the true value of n lies somewhere in the 
interval ${\rm n} = 1.2 \sim 1.4$ (hopefully, the planned new 
observational missions will determine this
index more accurately). Since ${\rm n} \equiv 2\beta+5$, this gives us the 
parameter $\beta$ in the
interval $\beta = -1.9 \sim -1.8$. In fact, 
the value ${\rm n} = 1.4$ $(\beta = - 1.8)$         
is the largest one for which the entire theoretical approch is well posed,
since this value requires the pumping field to be too strong: 
the Hubble radius at the end of the initial stage would have been 
only a little larger than $l_{Pl}$. With the adopted $A$ and $\beta$,
the frequency $\nu_c$ falls in the region $\nu_c = 10^{10} Hz$ or higher. An
allowed intermidiate stage of expansion governed by a ``stiff" matter [14]
can affect the spectral slope at frequencies somewhat lower than $\nu_c$, but
still outside the interval accessible to the existing detection techniques. 
The details of a short transition from the initial stage to the 
radiation-dominated 
stage are irrelevant for our discussion, since they can alter the 
signal only at frequencies around $\nu_c$. It is worth recalling [13] that
a confirmation of any ${\rm n} > 1$ ($\beta > -2$) would mean that the
very early Universe was not driven by a scalar field - the cornerstone of
inflationary considerations - because the ${\rm n} > 1$ ($\beta > -2$) 
requires the effective equation of state at the initial stage to be 
$\epsilon + p < 0$, but this cannot be accomodated by any scalar field with
whichever scalar field potential. Obiously, the available data do not
prove yet that ${\rm n} > 1$ but this possibility is not ruled out either.  
It is important to emphasize that the existing disagreement (see [11]) 
with regard to the validity of the inflationary prediction of infinitely 
large amplitudes of density perturbations for the part of the spectrum with
the spectral index ${\rm n} =1$ ($\beta = -2$) does not affect the assumptions
and conclusions of this paper which analyses the cases $\beta > - 2$. 
\par 
We switch now from cosmology to experimental predictions
in terms of laboratory frequencies $\nu$ and intervals of time $t$     
$(c{\rm d}t = a(\eta_R){\rm d}\eta)$. Formula (12) translates into
\begin{equation}
\label{13}
h(\nu,t) \approx 
10^{-7}\cos[2\pi\nu(t - t_{\nu})]\left(\frac{\nu}{\nu_H}\right)^{\beta+1},
\end{equation} 
where $\nu_H = 10^{-18} Hz$, and $t_{\nu}$ is a deterministic (not random) 
function of frequency which does not vary significantly 
on the intervals $\Delta\nu \approx \nu$. We take $\nu = 10^2 Hz$ as 
the representative frequency for the 
ground-based laser interferometers. The expected sensitivity of the 
initial instruments at $\nu = 10^2 Hz$ is
$h_{ex} = 10^{-21}$ or better. The theoretical 
prediction at this
frequency, following from (13), is $h_{th} = 10^{-23}$ for $\beta = -1.8$, 
and $h_{th} = 10^{-25}$ for $\beta = -1.9$. Therefore, the gap between
the signal and noise levels is from 2 to 4 orders of magnitude. This gap
should be covered by a sufficiently long observation time $\tau$.
The duration
$\tau$ depends on whether the signal has any temporal 
signature known in advance, or not.  
\par
It appears that the periodic structure (13) should survive in 
the instrumental window of sensitivity from $\nu_1$ (minimal frequency) 
to $\nu_2$ (maximal frequency). The mean square value of the field in this 
window is 
\begin{equation}
\label{14}
\int_{\nu_1}^{\nu_2}h^2(\nu,t)\frac{{\rm d}\nu}{\nu} =
10^{-14}\frac{1}{{\nu_H}^{2\beta+2}}\int_{\nu_1}^{\nu_2}\cos^2[2\pi\nu(t -
t_{\nu})]\nu^{2\beta+1}{\rm d}\nu~.
\end{equation}
Because of the strong dependence of the integrand on frequency, 
$\nu^{-2.6}$ or $\nu^{-2.8}$, the integral (14) is determined by 
its lower limit. This gives
\begin{equation}
\label{15}
\int_{\nu_1}^{\nu_2}h^2(\nu,t)\frac{{\rm d}\nu}{\nu} \approx
10^{-14}\left(\frac{\nu_1}{{\nu_H}}\right)^{2\beta+2}\cos^2[2\pi\nu_1(t -
t_1)]~.
\end{equation}
The explicit time dependence of the variance 
of the field, or, in other words, the explicit time dependence of 
the (zero-lag) temporal correlation function of the field, demonstrates 
that we are dealing with a non-stationary process (a consequence
of squeezing and severe reduction of the phase uncertainty). Apparently,
the search through the data should be based on the periodic 
structure at $\nu = \nu_1$.
\par
The response of an instrument to the incoming radiation is 	
$s(t) = F_{ij}h^{ij}$ where $F_{ij}$ depends on the position and 
orientation of the
instrument. The cross correlation of responses from two
instruments $\langle 0|s_1(t)s_2(t)|0\rangle$ will involve the overlap 
reduction
function [15 - 18], which we assume to be not much smaller than 1 [17]. 
The essential
part of the cross correlation will be determined by an expression of the
same form as (15).          
\par
The signal to noise ratio $S/N$ in the measurement of the amplitude 
of a signal with no specific known features increases as
$(\tau \nu_1)^{1/4}$. If 
the signal has known features exploited by the matched filtering technique, 
the $S/N$ increases as
$(\tau \nu_1)^{1/2}$. 
The guaranteed law $(\tau \nu_1)^{1/4}$ 
requires a reasonably short time $\tau = 10^6~{\rm sec}$ to improve the
$S/N$ by two orders of magnitude and to reach the level of 
the predicted signal with the extreme spectral index $\beta = -1.8$.
If the law $(\tau \nu_1)^{1/2}$ can be implemented,  
the same observation time $\tau = 10^6~{\rm sec}$ will allow the registration 
of the signal with the conservative spectral index $\beta = -1.9$. 
Even an intermediate law between
$(\tau \nu_1)^{1/4}$ and $(\tau \nu_1)^{1/2}$ 
may turn out to be sufficient. For the network of ground-based 
interferometers the expected $\nu_1$ is around
$30 Hz$, but we have used $\nu_1 = 10^2 Hz$ for a conservative 
estimate of $\tau$.
If the matched filtering technique can indeed be used, it can prove 
sufficient to have data from a single interferometer. 
\par
For the frequency
intervals covered by space intereferometers, solid-state detectors, 
and electromagnetic detectors, the
expected results follow from the same formula (13) and have been briefly
discussed elsewhere [13].    
\par
In conclusion,
the detection of relic (squeezed) gravitational waves may be awaiting 
only the first generation of sensitive instruments and an appropriate 
data processing strategy.   
\par
I appreciate useful discussions with S. Dhurandhar and
B. Sathyaprakash.
\par
-----------------------

\begin{itemize}
\item[[1]]
L. P. Grishchuk, Zh. Eksp. Teor. Fiz. {\bf 67}, 825 (1974)  
[JETP {\bf 40}, 409 (1975)]; Ann. NY Acad. Sci. {\bf 302}, 439 (1977). 
\item[[2]]
L. P. Grishchuk and Yu. V. Sidorov, Class. Quant. Grav.
{\bf 6}, L161 (1989); Phys. Rev. {\bf D42}, 3413 (1990).  
\item[[3]]
P. L. Knight, in {\it Quantum Fluctuations}, Eds. S. Reynaud, 
E. Giacobino, and J. Zinn-Justin, (Elsevier Science) 1997, p. 5. 
\item[[4]]
L. P. Grishchuk, in {\it Workshop on Squeezed States and
Uncertainty Relations}, NASA Conf. Publ. {\bf 3135}, 1992, p. 329;
Class. Quant. Grav. {\bf 10}, 2449 (1993).  
\item[[5]]
A. Abramovici {\it et. al.}, Science {\bf 256}, 325 (1992).  
\item[[6]]
C. Bradaschia {\it et. al.}, Nucl. Instrum. and Methods
{\bf A289}, 518 (1990).  
\item[[7]]
J. Hough and K. Danzmann {\it et. al.}, {GEO600 Proposal},
1994.  
\item[[8]]
W. Schleich and J. A. Wheeler, J. Opt. Soc. Am {\bf B4}, 1715 (1987); 
W. Schleich {\it et. al.}, Phys. Rev. {\bf A40}, 7405 (1989). 
\item[[9]] 
G. F. Smoot {\it et. al.}, Astroph. J. {\bf 396}, L1 (1992). 
\item[[10]] 
C. L. Bennet {\it et. al.}, Astroph. J. {\bf 464}, L1 (1996).
\item[[11]]
L. P. Grishchuk, Phys. Rev. {\bf D50}, 7154 (1994);
in {\it Current Topics in Astrofundamental Physics: Primordial Cosmology},
Eds. N. Sanchez and A. Zichichi, (Kluwer Academic) 1998, p. 539;
Report gr-qc/9801011.        
\item[[12]] 
A. A. Brukhanov {\it et. al.}, Report astro-ph/9512151. 
\item[[13]]
L. P. Grishchuk, Class. Quant. Grav. {\bf 14}, 1445 (1997). 
\item[[14]]
M. Giovannini, Phys. Rev. {\bf D58}, 1 September 1998 (to appear). 
\item[[15]]
P. F. Michelson, Mon. Not. R. astr. Soc. {\bf 227}, 933 (1987). 
\item[[16]]
N. L. Christensen, Phys. Rev. {\bf D46}, 5250 (1992). 
\item[[17]]
E. E. Flanagan, Phys. Rev. {\bf D48}, 2389 (1993). 
\item[[18]]
B. Allen, in {\it Relativistic Gravitation and Gravitational
Radiation}, Eds. J-A. Marck and J-P. Lasota (CUP, 1997) p. 373. 
\item[[19]]
M. Tegmark, Report astro-ph/9809201 
\end{itemize}
\end{document}